\documentclass{webofc}
\usepackage[varg]{txfonts}   
\usepackage{multicol}
\usepackage{xcolor}
\usepackage{hyperref}

\begin{document}
\title{Probing the evolution of galaxy clusters using SZ effect and non-thermal emission: first results from A1413}
%
%

\author{\lastname{M. Pachchigar}\inst{1}\fnsep\thanks{e-mail:mohini.pachchigar@vuw.ac.nz} \and
        \lastname{Y. Perrott}\inst{1} \and
        \lastname{T. Parashar}\inst{1} 
}

\institute{Victoria University of Wellington 
          }

\abstract{
 Mass is the most fundamental property of galaxy clusters. However, measuring it is still a challenge. Calibrating mass from intracluster medium observables such as the Sunyaev-Zel’dovich (SZ) effect is subject to uncertainty and biases because of the hydrostatic equilibrium assumption. On the other hand, merging cluster systems have been shown to exhibit radio emission which implies a link with disturbances from hydrostatic equilibrium. We present work on studying deviations of galaxy cluster gas pressure profile from the average (universal) pressure profile using an example of galaxy cluster Abell 1413  with SZ effect data from the Arcminute Microkelvin Imager and \emph{Planck}. This cluster has also been observed at low radio frequency with the Murchison Widefield Array allowing the investigation of links between gas pressure profile deviations and the presence of radio emission.
}
\maketitle

\section{Introduction}
\label{intro}

It is common to assume hydrostatic equilibrium equation (HSE) to derive cluster mass from thermodynamic properties of the ICM. However, the cluster pressure profile has two components: thermal pressure, and non-thermal pressure which induces deviations from HSE. The non-thermal pressure component becomes particularly significant for merging or disturbed clusters \citep{Pratt:2019}. SZ effect probes thermal pressure, but measuring non-thermal pressure is ambiguous. 
\\
The non-thermal electrons produce diffuse radio emission via synchrotron radiation which are broadly categorised as radio halos, radio relics and radio mini halos. Radio halos are Mpc size, centrally located, non-polarised emission. Relics are polarised, elongated, filament-like structures that are found at the cluster peripheries. Radio mini halos are similar to halos but smaller in size ($\sim$kpc) and they are observed in cool-core clusters \citep{vanWeeren:2019}. Mini halos are intriguing because their nature and formation mechanism are not well known. Currently only $\sim 30$ clusters hosting mini halos have been observed \citep{GendronMarsolais:2017}.
\\
We are currently studying a sample of nine galaxy clusters selected from the \emph{Planck} catalogue \cite{PlanckSZ} at signal-to-noise ratio (SNR) > 5 and with good detection in SZ by the Arcminute Microkelvin Imager (AMI) \cite{AMI}. These clusters also have Murchinson Widefield Array (MWA) \cite{Tingay:2013} observations which will enable us to investigate the link between thermal and non-thermal ICM properties. In this paper we present the first results from galaxy cluster A1413 which hosts a mini halo.

\section{Methodology/Data analysis}
\label{methodology}

\subsection{SZ data}

\noindent{\bf Observations:}
We use SZ data from AMI and $\emph{Planck}$. AMI operates over a frequency range of 13.5 -- 18.0 GHz. It is a pair of radio synthesis telescopes, the Small Array (SA) and the Large Array (LA) \citep{AMI}. The AMI LA has higher angular resolution of $\approx$ 30 arcsec and is insensitive to arcminute-scale SZ features. The AMI SA has angular resolution of $\approx$ 3 arcmin, and has the short baselines required for sensitivity to the arcminute-scale SZ. The LA is used to detect contaminating radio sources, which are modelled and removed from the SA view of the SZ as described below.
\\
We also use the High Frequency Instrument (HFI) data from the \emph{Planck} satellite which covers a range of frequency from $\approx$ 100 -- 900 GHz with beam sizes varying from $\approx$ 10 -- 5 arcmin \citep{PlanckCollab:2014}. AMI being an interferometer cannot capture the large extended emission in the cluster outskirts. However, \emph{Planck} can easily observe this extended emission and thus combining their observations is beneficial.


\noindent{\bf Model fitting:}
The \textit{Generalised Navarro Frenk White model} (GNFW) is commonly used to define cluster pressure profiles \cite{Nagai:2007}. Equation \ref{eq.GNFW} describes the GNFW profile where $P_{0}$ is the pressure normalisation factor. The various shape parameters are: $\gamma$ describing the inner slope ($r \ll r_{p}$), $\alpha$ describing the intermediate slope ($r = r_{p}$) and $\beta$ describing the outermost slope ($r \gg r_{p}$). $x=r/r_{p}=\theta/\theta_{s}$ where $r_{p}$ and $\theta_{s}$ are the characteristic physical and angular radius respectively. $r_{p} = \theta_{s}\times D_{A}$ where $D_{A}$ is the angular diameter distance.

\begin{equation}
    P(x)= \frac{P_{0}}{x^{\gamma}[1+x^{\alpha}]^{(\beta - \gamma)/\alpha}}
    \label{eq.GNFW}
\end{equation}

A comparison between AMI and \emph{Planck} data is possible using a model fitting algorithm for various parameters. We use the \textsc{McAdam} software to perform Bayesian analysis on the AMI data as in \cite{Perrott:2015}, constraining the cluster model parameters simultaneously with the radio source parameters. The fitting is done in the uv-plane, treating all radio sources as unresolved (a good approximation for the low-resolution SA). For $\emph{Planck}$ data, we use the \textsc{PowellSnakes} software \cite{Carvalho:2012} to perform a similar Bayesian analysis, and then combine the two analyses to produce joint posterior constraints \citep{Perrott:2019}. Our sampling parameters for the cluster pressure profile are $Y_{\mathrm{tot}}$ (total integrated Compton $y$ parameter), $\theta_{s}$, $\alpha$, $\beta$ and $\gamma$. The priors on these parameters are listed in table \ref{priors}.
\\

\begin{table}
\centering
\caption{Priors for evaluating cluster posteriors. $\mathcal{U}$ represents uniform prior and $\delta$ a delta function.}
\label{priors}       
\vspace*{-1mm}
\begin{tabular}{ll}
\hline
\textbf{Parameter} & \textbf{Prior distribution} \\\hline
$Y_{tot}$ & $\mathcal{U}$[0.0005 $\mathrm{arcmin^{2}}$, 0.5 $\mathrm{arcmin^{2}}$]  \\
$\theta_{s}$ & $\mathcal{U}$[1.3 $\mathrm{arcmin}$, 60 $\mathrm{arcmin}$]  \\
$\alpha$ & $\mathcal{U}[0.1, 3.5]$  \\
$\beta$ & $\mathcal{U}[3.5, 7.5]$  \\
$\gamma$ & $\delta$[0.43] \\\hline
\end{tabular}
\vspace*{0.5mm}  
\end{table}

\cite{Arnaud:2010} (hereafter A10) provided a hybrid average cluster pressure profile, combining X-ray observations and simulation for a sample of low redshift clusters. Their values of the shape parameters and normalised pressure are shown in table \ref{UPP}. Later, \cite{Ghirardini:2019} (hereafter G19) calculated an average GNFW profile from X-ray and SZ observations of the XCOP sample. Their best-fit values are listed in table \ref{UPP}. These average profile shapes will be compared to our AMI and \emph{Planck} constraints. We choose A10 as a comparison profile because it has been used extensively as a standard, universal profile in the literature. We choose G19 as another comparison profile since it has been calculated using a sample of clusters similar in mass and redshift to A1413, using SZ data to measure the outskirts unlike A10 which used simulations. More extensive comparisons will be made to other literature estimates of average profiles (eg \cite{Ghirardini:2017, McDonald:2014, Tramonte:2023, Sayers:2023}) in future work. Do note that we fix the value of $\gamma$, which our (low-resolution) data are relatively insensitive to, to the XCOP average value. The effect of this will also be investigated further in future work.

\begin{table}
\centering
\caption{Pressure profile parameters from \cite{Arnaud:2010} (A10) and \cite{Ghirardini:2019} (G19) }
\vspace*{-1mm}
\label{UPP} 
\begin{tabular}{lll}
\hline
\textbf{Parameter} & \textbf{A10} & \textbf{G19} \\\hline
$P_{0}$ & $8.403 h_{70}^{-3/2}$ & $5.68 \pm 1.77$  \\
$c_{500}$ & 1.177 & $1.49 \pm 0.30 $  \\
$\alpha$ & 1.0510 & 1.33  \\
$\beta$ & 5.4905 & $4.40 \pm 0.41 $  \\
$\gamma$ & 0.3081 & $0.43 \pm 0.10 $ \\\hline
\end{tabular}
\vspace*{0.5mm}  
\end{table}

\subsection{Synchrotron data}

\noindent {\bf Observation:} 
In order to observe the synchrotron emission, a low frequency radio telescope is necessary. We use data from the MWA to study the synchrotron emission. MWA has a wide frequency range from $\approx$ 80 -- 300 MHz, which is ideal to detect such diffuse radio emissions. It has an angular resolution of $\approx 1$ arcmin and captures all large angular scales relevant for our cluster sample \citep{Tingay:2013}.

\noindent{\bf Analysis:}
We use the GLEAMX pipeline \footnote{https://github.com/tjgalvin/GLEAM-X-pipeline} to make the MWA maps \citep{GLEAMX}. Future work will include calibration improvements to the MWA image and calculating the minihalo flux densities at the two MWA frequencies (154 MHz and 199 MHz).

\section{Results and Discussion}
\label{results}

We present our first results from this cluster sample analysis for the galaxy cluster A1413. Abell 1413 is a galaxy cluster at the redshift of $z=0.143$ and mass of $M_{500}=5.98_{-0.40}^{+0.48} \times 10^{14}$ $M_{\odot}$ \citep{PlanckCollab:2014}. The dynamic state of this cluster is not very clear in literature. There is evidence of A1413 being a relaxed cluster \cite{Vikhlinin:2005, Govoni:2009}, however \cite{Giacintucci:2017} classify it as an unrelaxed cluster. Recently, others \cite{Lusetti:2023, Riseley:2023} have classified A1413 as an intermediate cluster.

\subsection{Abell 1413}
\begin{figure}[h]
\centering
\includegraphics[scale=0.8]{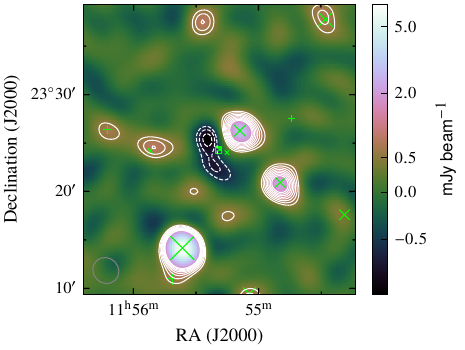}
\includegraphics[scale=0.8]{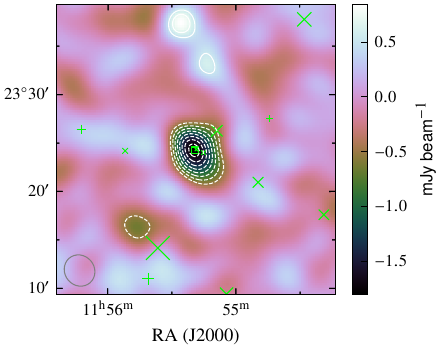}
\vspace*{-3mm}
\caption{AMI SA maps for cluster A1413. \textit{Left:} the AMI SA image with confusing radio point sources. \textit{Right:} the same AMI image after source subtraction. The green $\times$ and $+$ are the positions of the radio point sources observed with AMI LA and with their sizes scaled to flux density values. Central `$\square$' represents the position of cluster centre. The image is centered at the X-ray-derived position of the cluster. White contours are at $\pm$ 3--9 $\sigma$ level; dashed contours are negative. The telescope's synthesised beam is shown in grey on left bottom corner of each map. We achieve a high-significance detection of the SZ decrement at the cluster position after subtracting the confusing point sources.}
\label{A1413_AMI}       
\end{figure}

 The AMI SA map of the cluster is presented in figure \ref{A1413_AMI}, before and after source subtraction. The negative decrement of 9 $\sigma$ in the centre is a clear detection of the SZ effect from A1413. Compact sources have been subtracted using the flux density and spectral index values fit simultaneously with the cluster model parameters.

The core looks very busy with multiple radio sources. Recent studies with MeerKAT and LOFAR by \cite{Riseley:2023, Lusetti:2023} resolve the cluster core further. A counterpart of BCG was identified and investigated in MeerKAT studies \cite{Riseley:2023, Trehaeven:2023}. A head-tail radio galaxy at redshift z=0.144 was detected to the west of BCG. These two components are unresolved from the minihalo in lower-resolution (AMI, MWA) observations.

\begin{figure}[h]

\centering
\includegraphics[scale=0.6, trim = {1mm 3mm 1mm 1mm}, clip]{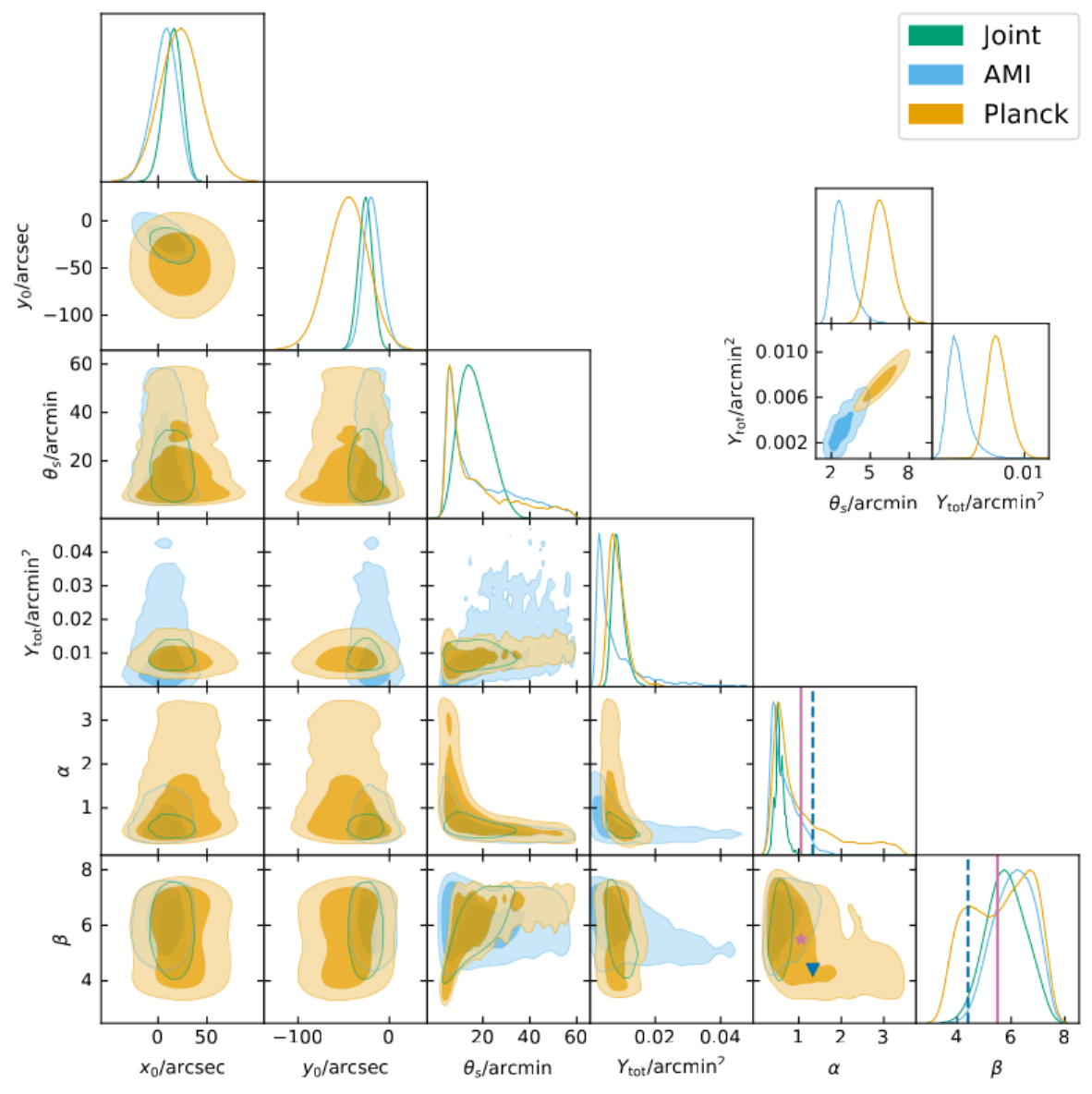}
\caption {\textit{Upper right:} $\theta_s$ and $Y_\mathrm{tot}$ constraints for individual datasets when the GNFW shape parameters are fixed to the G19 values. These incompatible constraints indicate that profile shape parameters are not appropriate for this cluster. Note that the posteriors are similarly incompatible when the GNFW shape parameters are fixed to A10 values. \textit{Lower:} Cluster parameter constraints with varying shape parameter $\alpha$ and $\beta$. Here, $x_{0}$ and $y_{0}$ are cluster position parameters. The magenta (star) and blue (triangle) marker in the $\alpha$ and $\beta$ posteriors represent fixed A10 and G19 values for corresponding parameters respectively.} 
\label{clus_post}       

\end{figure}

\noindent {\bf Cluster posteriors:} 
Figure \ref{clus_post} shows the posterior distributions for A1413 with AMI only (blue), \emph{Planck} only (orange) and AMI-\emph{Planck} joint (green) analysis using input priors as shown in table \ref{priors}. The upper right-hand plot showing only $\theta_s$ and $Y_\mathrm{tot}$, represents the constraints for the individual datasets when the GNFW shape parameters are fixed to the G19 values. The constraints are incompatible, which is an indication that the profile shape parameters are not appropriate for this cluster \cite{Perrott:2019}. The lower plot, including also the $\alpha$ and $\beta$ shape parameters, shows the corresponding results when these parameters are varied. It shows that the combination of AMI and \emph{Planck} data is effective in cutting down the degeneracy present in the GNFW model when only a limited range of angular scales is present in the measurement, producing constraints on $\alpha$ and $\beta$ which are different to both the A10 and G19 average values. This allows the $\theta_s$ and $Y_\mathrm{tot}$ values to come into agreement.

\begin{figure}[h]
\centering
\includegraphics[scale=0.60]{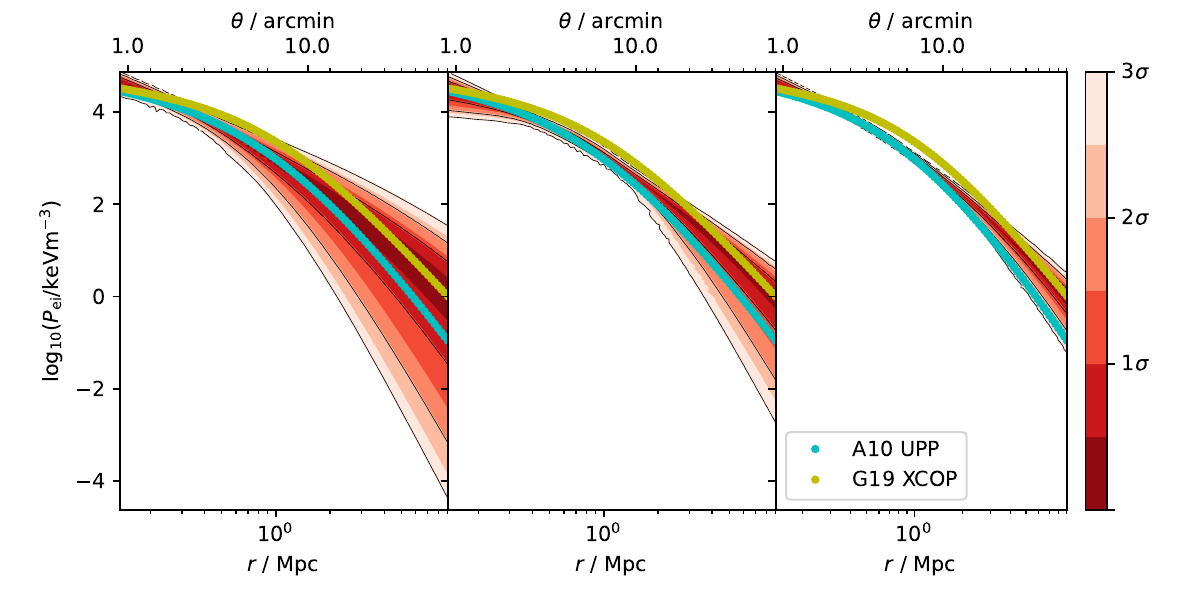}
\vspace*{-5mm}
\caption{\textit{Left:} AMI pressure profile, \textit{Middle:} \emph{Planck} pressure profile \textit{Right:} Joint analysis profile. Blue line represents A10 universal pressure profile. Green line corresponds to XCOP pressure profile derived by G19. All three profiles agree reasonably well with the A10 profile in the intermediate regions and with the G19 profile in outskirts. Note: we have fixed the interior shape parameter $\gamma$ to the G19 value.}
\label{pressure_profile}       

\end{figure}

\noindent {\bf Gas pressure profile:} 
From the above cluster parameter posteriors, we have derived pressure profile constraints for each analysis. These are shown in figure \ref{pressure_profile} in comparison with A10 and G19 profiles. All three profiles agree reasonably well with the A10 profile in the intermediate regions, but fall off less rapidly in the outskirts. In comparison, the G19 profile does not match in the intermediate regions but is a better match than the A10 profile in the outskirts. We note that we have fixed the interior shape parameter $\gamma$ to the G19 value. More work is required to understand whether the observed difference in our profile can be attributed to analysis methodology differences, X-ray/SZ pressure measurement differences, or could be tied to the dynamical state of the cluster.


\begin{figure}[h]
\centering
\sidecaption
\includegraphics[scale=0.8]{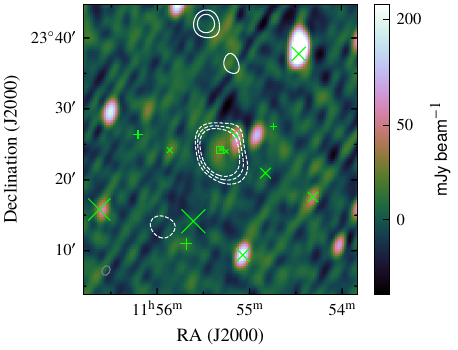}

\caption{ 154 MHz MWA image of cluster A1413 overlaid with source-subtracted AMI SA contours and point sources. The green $\times$ and $+$ are the positions of the radio point sources observed with AMI LA and with their sizes scaled to flux density values. Central `$\square$' represents the cluster centre. The image is centered at the X-ray-derived position of the cluster. White AMI SA contours are at $\pm $ 3--5 $\sigma$ level; dashed contours are negative. The telescope's synthesised beam is shown in grey on left bottom corner of the map. The extended central radio source is clearly visible. 
}
\label{A1413_MWA}   
\end{figure}

\noindent {\bf MWA results:} 
In the 154 MHz MWA image (figure \ref{A1413_MWA}) of the cluster A1413 an extended central radio source is clearly visible. Its brightness is the summation of flux densities of the mini-halo, BCG and head-tail radio galaxy. Some stripes are visible in the image from a nearby bright source indicating that further calibration  improvements are needed.
 

\section{Conclusion}
We present the results A1413, the first from our sample of galaxy clusters selected to have strong SZ detections with AMI and \emph{Planck}, and also low radio frequency observations with the MWA. Pressure profile constraints from this cluster show deviation from average cluster profiles in the literature, which may be linked to the dynamical state of this cluster. Our MWA observations detect the minihalo at 154MHz and 199 MHz.
 Next, we will investigate the spectral index value of the minihalo in A1413 using our observations.
 \\
 Future work on our cluster sample will include investigating the links between pressure profile shape, dynamical state and the presence or absence of synchrotron emission.

%
%
%
\vspace*{20mm}
\begin{multicols}{2}

\end{multicols}

\end{document}